\documentclass[preprint,prc,aps,showpacs,showkeys,groupedaddress,floatfix,superscriptaddress]{revtex4-1}
\usepackage{epsfig}
\usepackage{dcolumn}
\usepackage{bm}
\usepackage[utf8]{inputenc}
\usepackage{graphics}
\usepackage{graphicx}
\usepackage{epsfig}
\usepackage{amssymb}
\usepackage{amsmath}
\usepackage{color}
\usepackage{bbold}
\DeclareMathOperator\erf{erf}
\usepackage{subcaption}

\begin{document}

\title{Higher order GUP black hole based on COW experiment and Einstein-Bohr's photon box}

\author{Bilel Hamil}
\email{hamilbilel@gmail.com}
\affiliation{Département de TC de SNV, Université Hassiba Benbouali, Chlef, Algeria.}

\author{Houcine Aounallah}
\email{houcine.aounallah@univ-tebessa.dz}
\affiliation{Department of Science and Technology. Larbi Tebessi University, 12000 Tebessa, Algeria}

\author{Prabir Rudra}
\email{prudra.math@gmail.com}
\affiliation{Department of Mathematics, Asutosh College, Kolkata-700026, India}

\date{\today}
\begin{abstract}
In this work we have explored the effects of higher order generalized uncertainty principle (GUP), inspired from the quantum gravity COW experiment and the Einstein-Bohr's photon box thought experiment, on the properties of a black hole. Two different GUP models, namely, GUP to all orders in the Planck length model, and GUP with minimal length uncertainty and maximal momentum model are considered for our study. For each model, we have investigated the modified de Broglie formula, modification in gravitational phase shift, Einstein-Bohr's photon box, and the modified Hawking temperature and the tidal force of the GUP modified black hole.

\end{abstract}
\keywords{}
\pacs{}
\maketitle 

\section{Introduction}
Quantum theory of gravity has been one of the prime quest of theoretical physics for almost a century. There are various approaches to the theory, which ultimately predicts a minimum measurable length scale \cite{min1}. This along with the effects of generalized uncertainty principle (GUP) and modified dispersion relations (MDRs) on black hole (BH) thermodynamics have been the basis of the development of a quantum theory of gravity. Various theories of quantum gravity \cite{qg1, qg2, qg3} following a string theory approach catalyze the effects of GUP and MDRs. Other notable concepts paving way for the GUP are Loop quantum  \cite{lqg1}, BH gedanken exeriments \cite{gad1} and non-commutative spacetime algebra \cite{nc1, nc2, nc3}. These ideas basically show that the standard Heisenberg uncertainty principle needs generalization to incorporate the effects of quantum gravity.

The effects of GUP and MDRs on BH thermodynamics is intensively investigated in the literature \cite{bht1, bht2}. It has been seen that GUP prevents the complete evaporation of BH exactly as the Heisenberg uncertainty principle prevents the total collapse of hydrogen atom. As a result of this, at the end of the Hawking radiation process, an inert Planck size BH remnant exists which has zero entropy. This BH remnant is considered as a crucial clue towards understanding the concept of dark matter \cite{dm1, dm2} especially due to its inert nature. Special interest have been generated in computing the BH entropy and the sub-leading logarithmic correction \cite{lc1, lc2, lc3, lc4, lc5}. 

Gravity experiments with neutrons have been quite popular for decades. In 1975, Colella, Overhauser and Werner (COW) \cite{Collela} performed a ground breaking work where they demonstrated a phase shift induced by gravity. This experiment was performed on a neutron interferometer at the 2MW university of Michigan Reactor. In the reactor the neutron de-Broglie waves in the gravity potential are coherently split and separated. The final signal output is a result of the interference between these two splitted neutron waves. In order to test quantum theories of gravity, experiments where the outcome depends on both the gravitational acceleration $g$ and the Planck's constant $\hbar$ were designed. This helped to explore the effects of gravity and quantum interference using neutrons. These experiments involving the splitting of energy eigen states operate on the energy scale of pico-eVs, which can be used in measuring the fundamental constants \cite{cons1} and also in the quest of a non-Newtonian theory of gravity \cite{cons2}. Extensive reviews on the gravity experiments using neutrons as classical and quantum objects can be found in \cite{rev1, rev2}. 

The 1930 Solvay conference witnessed a famous debate between Einstein and Bohr, which is considered as a notable event in the development of quantum mechanics. According to Bohr \cite{bohr1} the argument revolved round the validity of the time-energy uncertainty relation. Einstein devised a thought-experiment, popularly known as the Einstein-Bohr's photon box, which could precisely predict the energy of the photon and its instant of arrival at the detector simultaneously, thus violating the energy-time uncertainly relation of quantum mechanics. The validity of this experiment would automatically imply the incompleteness of quantum mechanics and in turn the quantum-mechanical description of the physical reality will need serious revisions. Although this idea was satisfactorily countered by Bohr with his ingenious arguments, yet this thought-experiment remains a milestone in the gradual development of quantum mechanics. The concept of Einstein-Bohr's photon box finally led to a stronger idea known as the Einstein–Podolsky–Rosen paradox (EPR paradox) \cite{epr1}, where the three physicists argued that the picture of physical reality provided by quantum mechanics had missing links and was far from being complete.

Higher order corrections can be introduced in the standard uncertainty principles based on the COW experiment and the Einstein-Bohr's photon box thought-experiment. A quantum black hole (BH) based on these corrections will have interesting features. Motivated by this idea, we want to study such BHs and its properties in this work. The paper is organized as follows: Section II deals with the application of GUP to all orders in the Planck length model. In section III we study the GUP with minimal length uncertainty and maximal momentum model. Finally the paper ends with a conclusion in section IV.

\section{GUP to all orders in the Planck length model}
In this section we will apply a GUP formalism to all orders in the Planck length model. We will explore the modified de-Broglie formula, the modification in the gravitational phase shift, Einstein-Bohr's photon box and the modification in the Hawking temperature and the tidal force.

\subsection{Modified de Broglie formula}

Based on some theoretical considerations and for including gravitation with
quantum theory, the authors in \cite{Khireddine,Nouicer} proposed a new
adequate GUP to all orders in the Planck length, such that 
\begin{equation}
\left[ X,P\right] =i\hbar e^{\frac{\alpha ^{2}L_{p}^{2}}{\hbar ^{2}}%
P^{2}}=i\hbar \xi ,  \label{a}
\end{equation}%
\begin{equation}
\Delta X\Delta P=\frac{\hbar }{2}e^{-\frac{1}{2}W\left( -\frac{\alpha
^{2}L_{p}^{2}}{2\left( \Delta X\right) ^{2}}\right) },  \label{b}
\end{equation}

where $\xi =e^{\frac{\alpha ^{2}L_{p}^{2}}{\hbar ^{2}}P^{2}}$, $W\left(
u\right) $ is the Lambert function, $L_{p}$ is Planck length, $\alpha $ is a dimensionless number of order of unity and $\hbar$ is the reduced Planck's constant. The algebra defined by Eq. (\ref%
{b}) leads to a minimal uncertainty in position given by%
\begin{equation}
\left( \Delta X\right) _{\min }=\sqrt{\frac{e\alpha }{2}}L_{p}.
\end{equation}%
A representation of the commutation relations (\ref{a}) can be derived from
operators $\hat{x}$ and $\hat{p}$, verifying the usual commutation
relations, across the modification%
\begin{equation}
P=\hat{p};\text{ \  \  \ }X=i\hbar e^{\frac{\alpha ^{2}L_{p}^{2}}{\hbar ^{2}}%
p^{2}}\frac{d}{dp}.
\end{equation}%
Now, let us focus on the computation of the eigenvalue equation of the
position operator in momentum space. The spectrum of the position operator
takes, on the momentum space, the form of the differential equation,%
\begin{equation}
X\psi \left( p\right) =i\hbar e^{\frac{\alpha ^{2}L_{p}^{2}}{\hbar ^{2}}%
p^{2}}\frac{d}{dp}\psi \left( p\right) =x\psi \left( p\right) ,
\end{equation}

with solution%
\begin{equation}
\psi \left( p\right) =\psi _{x}\left( 0\right) \exp \left[ -\frac{ix\sqrt{%
\pi}}{2\alpha L_{p}}\erf{\left(\frac{\alpha L_{p}}{\hbar}p\right)} %
\right] .
\end{equation}%
The modified de Broglie wavelength can be obtained by a resemblance with
plane wave function $\psi \simeq e^{-\frac{i2\pi x}{\lambda }}$, accordingly
the modified de Broglie formula is given as

\begin{equation}
\lambda =\frac{4\sqrt{\pi }\alpha L_{p}}{\erf{\left(\frac{\alpha L_{p}}{\hbar}p\right)}},  \label{c}
\end{equation}%
where $\erf{\left( u\right)} $ is Gauss error function. Expand Eq.(\ref%
{c}) in the parameter $\alpha $, we have%
\begin{equation}
\lambda \approx \frac{2\pi \hbar }{p}+\frac{2\pi \hbar }{3}\left( \frac{%
\alpha L_{p}}{\hbar }\right) ^{2}p+\frac{2\pi \hbar }{90}\left( \frac{\alpha
L_{p}}{\hbar }\right) ^{4}p^{3}+...
\end{equation}

It becomes clear that as $\alpha \rightarrow 0$ we re-establish the ordinary
Broglie wavelength of the free particle. The modified de Broglie wavelength (%
\ref{c}) satisfies the following relation%
\begin{equation}
\frac{d}{dp}\left( \frac{2\pi }{\lambda }\right) =\hbar ^{-1}e^{-\frac{%
\alpha ^{2}L_{p}^{2}}{\hbar ^{2}}P^{2}}.
\end{equation}%
With the aim to make clear this equation, we construct a commutator in the
form%
\begin{equation}
\left[ X,\hat{k}\right] =i,  \label{d}
\end{equation}

where $\hat{k}=k\left( P\right) $.Thus, the commutation relation (\ref{a}),
can be written as

\begin{equation}
\left[ X,P\right] =\left[ X,\hat{k}\right] \frac{dP}{d\hat{k}}=i\frac{dP}{d%
\hat{k}}.  \label{com}
\end{equation}%
Comparing (\ref{a}) with (\ref{d}), we have%
\begin{equation}
\frac{d\hat{k}}{dP}=\hbar ^{-1}\xi ^{-1},
\end{equation}%
\begin{equation}
k\left( P\right) =\hbar ^{-1}\int e^{-\frac{\alpha ^{2}L_{p}^{2}}{\hbar ^{2}}%
P^{2}}dP.
\end{equation}%
It's clear that $\left[ P,k\right] =0$, which means that there is
eigenvector of eigenvalue $p$, which verify%
\begin{equation}
P\psi _{p}=p\psi _{p}\text{ \  \ and \ }\hat{k}\left( P\right) \psi
_{p}=k\left( p\right) \psi _{p}.  \label{e}
\end{equation}

If we compare Eq.(\ref{e}) with the usual commutation relation, we conclude
that $\hat{k}=-i\partial _{x}$. Thus, Eq.(\ref{e}) convert to%
\begin{equation}
-i\partial _{x}\psi _{p}=k\left( p\right) \psi _{p},
\end{equation}%
with solution $\psi _{p}=e^{ikx}$, with wavelength $\lambda =\frac{2\pi }{k}$%
. \ For this reason, \ $\hat{k}\left( P\right) $ introduced in (\ref{com})
can be seen as the wave-vector operator.

\subsection{Modification in gravitational Phase Shift}

It is now several decades since the Colella-Overhauser and Werner (COW)
experiments \cite{Overhauser,Collela,Werner} were performed. The authors
demonstrated, in principle, the first link between general relativity and
quantum mechanics. Werner and his co-workers, studied the the gravity
potential employing classical mechanics, but the equation of the
gravitational phase shift, observed in the experiments, includes both
Newton's constant of gravitation $G$ and Planck's constant $\hbar $. The
phase shift is given by

\begin{equation}
\Delta \varphi =\frac{mg}{\hbar v}A,  \label{phase}
\end{equation}%
where $g$ is the local acceleration due to gravity, $v$ denote the average
speed of neutrons, and $A$ the area enclosed by two interfering neutron
beams that propagate on two paths on a plane. Now, by considering the phase
shift of \cite{Xiang}, we have

\begin{eqnarray}
\Delta \varphi ^{\prime } &=&2\pi \left( \frac{L}{\lambda _{2}}-\frac{L}{%
\lambda _{1}}\right) ,  \notag \\
&=&2\pi L\left( \hat{k}_{2}-\hat{k}_{1}\right) =2\pi L\Delta \hat{k},  \notag
\\
&=&2\pi L\frac{\Delta \hat{k}}{\Delta P}\Delta P.
\end{eqnarray}%
where $\Delta P=P_{2}-P_{1}$, and for small value of $\Delta P$, we have 
\begin{equation}
\Delta \varphi ^{\prime }=2\pi L\xi ^{-1}\Delta P.  \label{f}
\end{equation}%
The neutron beams obey energy conservation law, so we have%
\begin{equation}
mgy=\frac{P_{2}^{2}-P_{1}^{2}}{2m}=v\Delta P.  \label{i}
\end{equation}%
By inserting Eq. (\ref{i}) into (\ref{f}), we have%
\begin{eqnarray}
\Delta \varphi ^{\prime } &=&\frac{mgyL}{\hbar vz}=\frac{mgA}{\hbar v\xi }, 
\notag \\
&=&\frac{m\bar{g}A}{\hbar v}.  \label{g}
\end{eqnarray}%
We obtained the modified phase shift from GUP to all orders in the Planck
length, where $\bar{g}=\frac{g}{\xi }$ \ and $A=yL$. When $\xi =1$, Eq. (\ref%
{g}) reduces to the ordinary form of Eq. (\ref{phase}). We remark that the
effect of GUP to all orders in the Planck length on COW experiment is
identical to the case that two neutron beams propagate in a gravitational
field described by the effective field strength $\bar{g}$.

\subsection{Einstein-Bohr's photon box}

Einstein proposed a gedanken experiment to evaluate the weight of the photon
in an effort to show the inconsistency of quantum mechanics \cite{Aharonov}.
He assumed that there is a box containing photon gas, with a completely
reflective wall and suspended using a spring scale. Inside the box there
is a perfect clock mechanism that can open and close the shutter. Following
the strategy of Einstein, when a photon was emitted from the box, the ideal
clock can measure the emission time interval exactly (i.e. $\Delta
t\rightarrow 0$). The energy of the emitted photon can be estimated by
establishing the difference in the mass of the box. From Einstein's outlook,
the time interval necessary for the photon to radiate exactly $\Delta
t\rightarrow 0$. Consequently, $\Delta t\Delta E\rightarrow 0$, which break
the Heisenberg uncertainty principle. From another point of view, based on
Einstein's general theory of relativity, Bohr mentioned that Einstein did
not take into account the consequence of time dilation caused by the
difference in gravitational potential, which conducts to an invalid
suggestion. One can confirm the validity of HUP as follows: during the
emission of the photon, the uncertainty in the momentum of the box is \cite%
{Aharonov}, 
\begin{equation}
\Delta p<gt\Delta m,  \label{k}
\end{equation}%
we employ (\ref{k}) in the HUP, we get

\begin{equation}
\hbar \leq \Delta x\Delta p<gt\Delta m\Delta x.
\end{equation}%
Applying $\Delta E=c^{2}\Delta m$, we obtain%
\begin{equation}
\Delta E>\frac{c^{2}\hbar }{gt\Delta x}.  \label{l}
\end{equation}

Following from general relativity, a clock in a gravitational potential
ticks slowly than a clock in free fall. Due to gravitational potential, two
clocks at different heights on top of the Earth will run at different rates.
So, if the difference in altitude is $\Delta x$, the rate of time $\frac{%
\Delta t}{t}$ is then \ 
\begin{equation}
\frac{\Delta t}{t}=\frac{g\Delta x}{c^{2}}.  \label{m}
\end{equation}%
Substituting eq. (\ref{m}) into eq. (\ref{l}), we find the uncertainty
relation%
\begin{equation}
\Delta t\Delta E>\hbar .
\end{equation}

In what follows, we consider Bohr's argument throughout the GUP to all
orders in the Planck length and we will determine the relationship between
the weight of the photon in the gedaken experiment and the corresponding
quantities in quantum mechanics. We follow the same method that was used above. We simply find

\begin{equation}
\frac{\Delta p_{\min }}{t}=\frac{\xi \hbar }{t\Delta X}\leq g\Delta m,
\end{equation}

or 
\begin{equation}
\Delta x\Delta p_{\min }=\xi \hbar \leq t\Delta m\left( g\Delta X\right) .
\label{n}
\end{equation}

Now multiplying \ Eq. (\ref{n}) by $c^{2}$ and using the relation $\Delta
E=c^{2}\Delta m$, we get%
\begin{equation}
\Delta E\Delta t\geq \frac{\xi \hbar c^{2}}{g\Delta x}=\frac{\hbar c^{2}}{%
\bar{g}\Delta X},  \label{o}
\end{equation}

The difference among (\ref{l}) and (\ref{o}) is that $g$ is replaced by%
\begin{equation}
\bar{g}=\frac{g}{\xi }=ge^{-\frac{\alpha ^{2}L_{p}^{2}}{\hbar ^{2}}p^{2}}.
\end{equation}%
In the limit $\alpha \rightarrow 0$, we obtain the HUP limit. Next, as done
in \cite{Xiang,Mohamed,Nasrin,Hassan}, we can considered the change in the
acceleration of gravity as a correction to the mass of the particle, such
that%
\begin{equation}
\bar{g}=\frac{g}{\xi }=\frac{GM}{\xi r^{2}}=\frac{G\bar{M}}{r^{2}},
\end{equation}

where 
\begin{equation}
\bar{M}=\frac{M}{\xi },  \label{eff}
\end{equation}
is the effective black hole mass and $G$ is the gravitational constant. Now, we replace $M$ by $\bar{M}$ we obtain
a modified Schwarzschild metric as follows%
\begin{equation}
ds^{2}=-\left( 1-\frac{2G\bar{M}}{r^{2}}\right) dt^{2}+\left( 1-\frac{2G\bar{%
M}}{r^{2}}\right) ^{-1}dr^{2}+r^{2}d\Omega .
\end{equation}

This metric is characterized by its dependency on different scales of
momenta and is similar to the gravity's rainbow \cite{Magueijo}.

\subsection{Modified Hawking temperature and the tidal force}

\begin{figure}[ht]
\begin{center}
\includegraphics[width=0.7\textwidth]{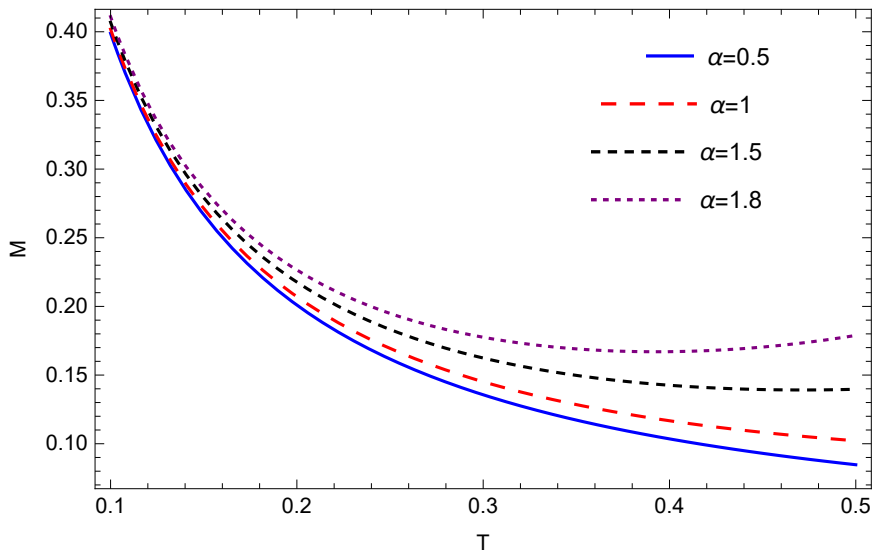}
\end{center}
\vspace{-0.5cm}
\caption{Plot for the black hole mass $M$ as a function of temperature $T$ for different values of $\alpha$. This is a graphical representation and the other constants are chosen to be unity. Plot not to scale.}
\label{figf}
\end{figure}

In the standard Hawking picture, the temperature of the Schwarzschild black
hole of mass $M$ is \cite{Hawking},%
\begin{equation}
T_{H}=\frac{\hbar }{8\pi K_{B}GM}.  \label{Hawmass}
\end{equation}%
where $K_B$ is the Boltzmann constant. To examine the effects of GUP to all orders in the Planck length on black
hole temperature we replace the black hole mass $M$ by the effective mass $%
\bar{M}$ and we assume $p=K_{B}T$, we find

\begin{equation}
M=\frac{\hbar }{8\pi GK_{B}T}e^{\frac{\alpha ^{2}L_{p}^{2}}{\hbar ^{2}}%
K_{B}^{2}T^{2}},  \label{p}
\end{equation}%
and for a small value of $\alpha $, the black hole mass can be extended as%
\begin{equation}
M=\frac{\hbar }{8\pi GK_{B}T}\left[ \allowbreak 1+\frac{\alpha ^{2}L_{p}^{2}%
}{\hbar ^{2}}K_{B}^{2}T^{2}+\frac{1}{2}\left( \frac{\alpha ^{2}L_{p}^{2}}{%
\hbar ^{2}}\right) ^{2}K_{B}^{4}T^{4}+...\allowbreak \right] ,  \label{mass}
\end{equation}

which reduces to Hawking's formula, $M=\frac{\hbar }{8\pi GK_{B}T}$, for $%
\alpha \rightarrow 0$. The expression (\ref{mass}) demonstrates that for
larger temperatures $T>$ $\frac{\hbar }{K_{B}\alpha L_{p}}$, the black hole
mass augments with temperature. In this scenario, such a behavior is
prohibited by the cut-off caused by GUP. In Fig.(\ref{figf}) we have plotted the black hole mass $M$ against the temperature $T$. It is seen that the mass gradually decays with temperature and settles to a constant value corresponding to the remnant mass. It is interesting to transpose Eq. (%
\ref{p}) and write the temperature $T$ of the black hole as a function of the mass $%
M$,

\begin{equation}
T=\frac{1}{8\pi L_{p}^{2}M}e^{-\frac{1}{2}W\left( -\frac{1}{e}\left( \frac{%
M_{0}}{M}\right) ^{2}\right) }.
\end{equation}

where $M_{0}=\sqrt{\frac{e}{2}}\frac{M_{p}}{2}.$

\vspace{1cm}

We now move to calculate the specific heat $C$ of the black hole. The heat
capacity of the black hole can be defined as%
\begin{equation}
C=\frac{dM}{dT}.  \label{def}
\end{equation}%
After some straightforward algebra we find 
\begin{equation}
C=-\frac{\hbar }{8\pi GK_{B}T^{2}}\left[ 1-\frac{2\alpha ^{2}L_{p}^{2}}{%
\hbar ^{2}}K_{B}^{2}T^{2}\right] e^{\frac{\alpha ^{2}L_{p}^{2}}{\hbar ^{2}}%
K_{B}^{2}T^{2}}.  \label{heat}
\end{equation}%
From Eq.(\ref{heat}), one can see that there exists a temperature at which
the heat capacity vanishes. The radiation process stops at this temperature
of the black hole with a finite mass termed as the remnant mass. Setting $%
C=0 $, this yield%
\begin{equation}
T_{rem}=\frac{\hbar }{\sqrt{2}K_{B}\alpha L_{p}}.  \label{trem}
\end{equation}%
We substitute Eq. (\ref{trem}) in \ Eq. (\ref{p}), we find the expression of
the remnant mass%
\begin{equation}
M_{rem}=\frac{\sqrt{2e}\alpha L_{p}}{8\pi G}.
\end{equation}%

\begin{figure}[ht]
\begin{center}
\includegraphics[width=0.7\textwidth]{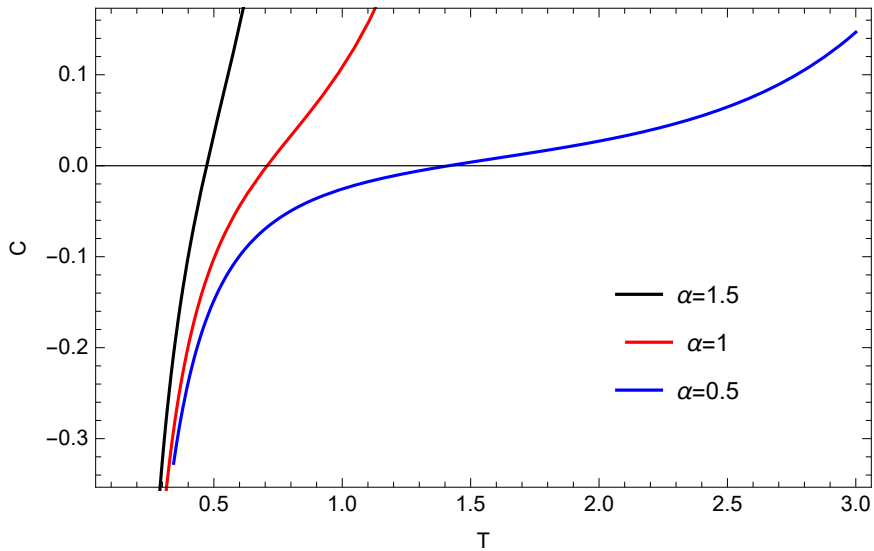}
\end{center}
\vspace{-0.5cm}
\caption{Plot for the specific heat capacity $C$ as a function of temperature $T$ for different values of $\alpha$. This is a graphical representation and the other constants are chosen to be unity. Plot not to scale.}
\label{figse}
\end{figure}

In Fig.(\ref{figse}) we have plotted the specific heat capacity $C$ against the temperature $T$. It is seen that the heat capacity undergoes a transition from negative to positive value. The temperature at which the transition occurs is the remnant temperature $T_{rem}$ corresponding to $C=0$. On the other side, for two virtual particles with energy $\Delta E$ separated by
a distance $\Delta X$, the tidal force $F$ reads as%
\begin{equation}
F=\frac{2GM}{r^{3}}\Delta E\Delta X.
\end{equation}%
So, the momentum uncertainty due to the tidal force is 
\begin{equation}
\Delta P=F\Delta t=\frac{2GM}{r^{3}}\Delta t\Delta E\Delta X,
\end{equation}%
where $\Delta t$ is the life time of the particle. Multiplying both side of
the above inequality by $\left( \Delta P\right) $ and using the Heisenberg
uncertainty relationships, we get

\begin{equation}
\left( \Delta P\right) \simeq \sqrt{\frac{\hbar ^{2}GM}{r^{3}}}.  \label{q}
\end{equation}%
substituting (\ref{q}) into (\ref{eff}), the effective black hole mass can
be rewritten as%
\begin{equation*}
\bar{M}=Me^{-\frac{\alpha ^{2}L_{p}^{2}}{r^{3}}GM}
\end{equation*}

\section{GUP with minimal length uncertainty and maximal \ momentum model}
Here we will explore the GUP effects with the minimal length uncertainty and maximal momentum model. In this case also we will investigate the same four scenarios as done in the previous section.

\subsection{Modified de Broglie formula}

In order to investigate the concept of minimal length uncertainty and
maximal \ momentum into quantum mechanics, we have considered the following
form of deformed Heisenberg relations \cite{Pedram2012}%
\begin{equation}
\left[ X,P\right] =\frac{i\hbar }{1-\beta P^{2}}=i\hbar z,
\end{equation}%
\begin{equation}
\Delta X\Delta P=\frac{\frac{\hbar }{2}}{1-\beta \left( \Delta P\right) ^{2}}%
.  \label{PED}
\end{equation}%
where $z=\frac{1}{1-\beta P^{2}}$, $\beta $ is the GUP parameter. This
algebra leads to a minimal uncertainty in position and and maximal \
momentum given by%
\begin{equation}
\left( \Delta X\right) _{\min }=\frac{3\sqrt{3}}{4}\hbar \sqrt{\beta };\text{
\  \  \ }P_{\max }=\frac{1}{\sqrt{\beta }}.
\end{equation}

we consider 
\begin{equation}
P=p;\text{ \ }X=\frac{i\hbar }{1-\beta p^{2}}\frac{d}{dp}
\end{equation}%
The eigenvalue problem for the position operator 
\begin{equation}
X\psi \left( p\right) =\frac{i\hbar }{1-\beta p^{2}}\frac{d}{dp}\psi \left(
p\right) =x\psi \left( p\right)
\end{equation}%
we have the solution 
\begin{equation}
\psi \left( p\right) =\psi \left( 0\right) exp\left( -\frac{ixp}{\hbar }%
\left( 1-\frac{\beta }{3}p^{2}\right) \right)
\end{equation}%
de Broglie formula is obtained by comparing the momentum eigenstate with a
plane wave function $exp\left( -\frac{2\pi ix}{\lambda }\right) $ the
modified de Broglie relation is given by 
\begin{equation}
\lambda =\frac{2\pi \hbar }{p\left( 1-\frac{\beta }{3}p^{2}\right) }
\end{equation}%
It becomes clear that as $\beta \rightarrow 0$ we re-establish the ordinary
Broglie wavelength of the free particle $\lambda =\frac{2\pi \hbar }{p}$.
and The modied de Broglie wavelength satises the following relation 
\begin{equation}
\frac{d}{dp}\left( \frac{2\pi }{\lambda }\right) =\hbar ^{-1}z^{-1}
\end{equation}%
In order to explain this formula, we first construct a commutator as follows 
\begin{equation}
\left[ X,\hat{k}\right] =i
\end{equation}%
where $\hat{k}=k\left( p\right) $ .Thus, the commutation relation (44), can
be written as 
\begin{equation}
\left[ X,P\right] =\left[ X,\hat{k}\right] \frac{dP}{d\hat{k}}=i\frac{dP}{d%
\hat{k}}
\end{equation}%
Comparing (44) with (10), we have 
\begin{equation}
\frac{d\hat{k}}{dp}=\hbar ^{-1}z^{-1}
\end{equation}%
and 
\begin{equation}
k\left( p\right) =\hbar ^{-1}\int \left( 1-\beta p^{2}\right) dP=\frac{p}{%
\hbar }\left( 1-\frac{\beta }{3}p^{2}\right)
\end{equation}%
Its clear that $\left[ P,k\right] =0$.

\subsection{Modification in gravitational Phase Shift}

We follow the same arguments of the previous sections and express

\begin{equation}
\Delta \varphi ^{\prime }=2\pi \left( \frac{L}{\lambda _{2}}-\frac{L}{%
\lambda _{1}}\right) =2\pi L\left( \hat{k}_{2}-\hat{k}_{1}\right) =2\pi L%
\frac{\Delta \hat{k}}{\Delta P}\Delta P
\end{equation}

\begin{equation}
\Delta \varphi ^{\prime }=2\pi Lz^{-1}\Delta P
\end{equation}%
using the energy conservation law Eq.(19), we obtain 
\begin{equation}
\Delta \varphi ^{\prime }=\frac{mgyL}{\hbar \upsilon }z^{-1}=\frac{%
mg^{\prime }A}{\hbar \upsilon }
\end{equation}%
where 
\begin{equation}
g^{\prime }=\frac{g}{z}
\end{equation}%
When $\beta \rightarrow 0$ so $g^{\prime }=g$ we returns to the earlier
result predicted by usual quantum theory.

\subsection{Einstein-Bohr's photon box}

\begin{equation}
\Delta x\Delta p_{min}=\hbar z
\end{equation}
where 
\begin{equation}
\Delta p_{min}\leq gt\Delta m
\end{equation}
and

\begin{equation}
\Delta x\Delta p_{min}=\hbar z\leq gt\Delta m\Delta x
\end{equation}%
or 
\begin{equation}
\frac{\Delta p_{min}}{t}=\frac{\hbar z}{t\Delta x}\leq g\Delta m
\end{equation}%
multiplying Eq. (27) by $c^{2}$ 
\begin{equation}
\hbar zc^{2}\leq gt\left( c^{2}\Delta m\right) \Delta x
\end{equation}%
where $\Delta E=c^{2}\Delta m$ we have 
\begin{equation}
\Delta E\geq \frac{\hbar zc^{2}}{gt\Delta x}=\frac{\hbar c^{2}}{g^{\prime
}t\Delta x}
\end{equation}%
where 
\begin{equation}
g^{\prime }=\frac{g}{z}=g\left( 1-\beta p^{2}\right)
\end{equation}%
Note that when $\beta =0$, we find the HUP limit. Then, we express the
effective mass from the definition

\begin{equation}
g^{\prime }=\frac{g}{z}=\frac{GM}{zr^{2}}=\frac{G\bar{M}}{r^{2}}
\end{equation}%
where 
\begin{equation}
\bar{M}=\frac{M}{z}=M\left( 1-\beta p^{2}\right)  \label{GUPM}
\end{equation}%
Next, as done in previous sections , we change $M$ with $\bar{M}$ and
examine a new Schwarzschild metric with the form

\begin{equation}
ds^{2}=-\left( 1-\frac{2G\bar{M}}{r^{2}}\right) dt^{2}+\left( 1-\frac{2G\bar{%
M}}{r^{2}}\right) ^{-1}dr^{2}+r^{2}d\Omega .
\end{equation}

\subsection{Modified Hawking temperature and the tidal force}

Let us then examine the effects of minimal length uncertainty and maximal
momentum brought by the GUP defined by (\ref{PED}), on the Hawking
temperature. Following the same procedure using in previous sections, we can
express the black hole temperature as follows:

\begin{equation}
T_{H}=\frac{\hbar }{8\pi K_{B}G\bar{M}}.
\end{equation}%
On the other hand, we identified $p$ with $K_{B}T$. \ Thus, we find the the
GUP-corrected Hawking mass 
\begin{equation}
M=\frac{\hbar }{8\pi K_{B}GT\left( 1-\beta K_{B}{}^{2}T^{2}\right) }.
\label{GUPmass}
\end{equation}

\begin{figure}[ht]
\begin{center}
\includegraphics[width=0.7\textwidth]{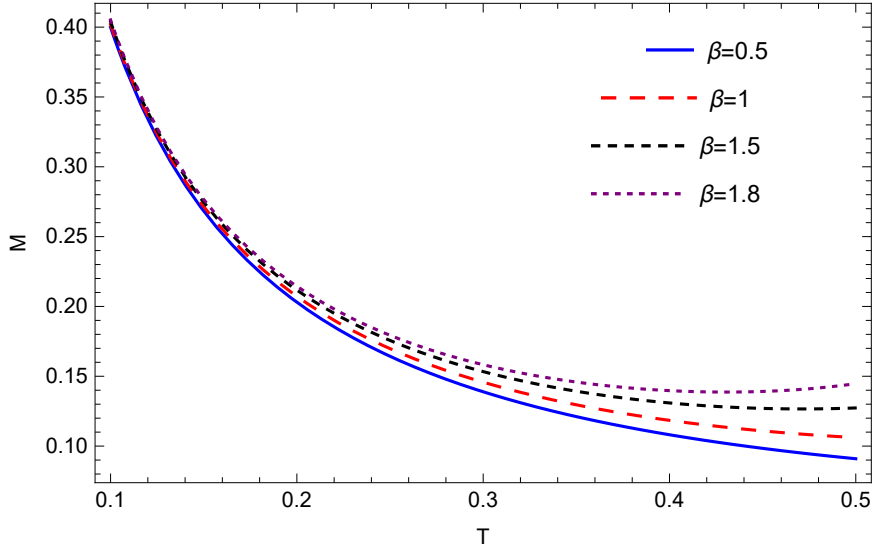}
\end{center}
\vspace{-0.5cm}
\caption{Plot for the mass $M$ as a function of temperature $T$ for different values of $\beta$. This is a graphical representation and the other constants are chosen to be unity. Plot not to scale.}
\label{figth}
\end{figure}

From this expression we see that the BH mass is only defined for ${}T\leq
T_{\max }=\frac{1}{K_{B}\sqrt{\beta }}$, and for a BH with a temperature
equal to $T_{\max }$, the Hawking mass tend to infinity. In Fig.(\ref{figth}) we have plotted the black hole mass $M$ against the temperature $T$. It is seen that the mass gradually decays with temperature and settles to a constant value corresponding to the remnant mass, as was the case in the previous section. The corrections to the standard Hawking mass are obtained by expanding eqn.(\ref{GUPmass}) in terms
of $\beta $:%
\begin{equation}
M=\frac{\hbar }{8\pi K_{B}GT}\left[ \allowbreak 1+\beta K_{B}^{2}T^{2}+\beta
^{2}K_{B}^{4}T^{4}+...\right] .
\end{equation}

Now , we can express the temperature in terms of the mass as

{\small 
\begin{eqnarray}
T &=&\frac{1}{3}\bigg \{ \frac{\left( \frac{1}{3}\right) ^{\frac{1}{3}%
}\left( \sqrt{3}T_{\max }\right) ^{2}}{\left[ -\sqrt{3}\left( \frac{M_{\min }%
}{M}\right) T_{\max }+\sqrt{3}T_{\max }\left( \frac{M_{\min }}{M}\right) 
\sqrt{1-4\left( \frac{M}{M_{\min }}\right) ^{2}T_{\max }^{4}}\right] ^{1/3}}+
\notag \\
&&\frac{\left[ -\sqrt{3}T_{\max }\left( \frac{M_{\min }}{M}\right) +\sqrt{3}%
T_{\max }\left( \frac{M_{\min }}{M}\right) \sqrt{1-4\left( \frac{M}{M_{\min }%
}\right) ^{2}T_{\max }^{4}}\right] ^{1/3}}{\left( \frac{1}{3}\right) ^{\frac{%
1}{3}}}\bigg \}.  \label{GUPT}
\end{eqnarray}%
} where%
\begin{equation}
M_{\min }=\frac{3\sqrt{3}}{16\pi }\frac{\hbar \sqrt{\beta }}{G},
\end{equation}%
is the black hole minimum mass. Eq.(\ref{GUPT}) readily implies the
existence of a critical mass above which the temperature will be a complex
quantity%
\begin{equation}
M\leq \frac{M_{\min }}{2T_{\max }^{2}}.
\end{equation}

\bigskip

Next, we derive the heat capacity function in the latter assumption. To this
end we employ Eq. (\ref{def}) and after straightforward algebra we find 
\begin{equation}
C=-\frac{\hbar \left( 1-3\beta K_{B}^{2}T^{2}\right) }{8\pi
K_{B}GT^{2}\left( 1-\beta K_{B}^{2}T^{2}\right) ^{2}}.
\end{equation}

To get the remnant mass, we set $C=0$ and this leads to

\begin{equation}
T_{rem}=\frac{T_{\max }}{\sqrt{3}};\text{ \  \ }M_{rem}=M_{\min }.
\end{equation}

\begin{figure}[ht]
\begin{center}
\includegraphics[width=0.7\textwidth]{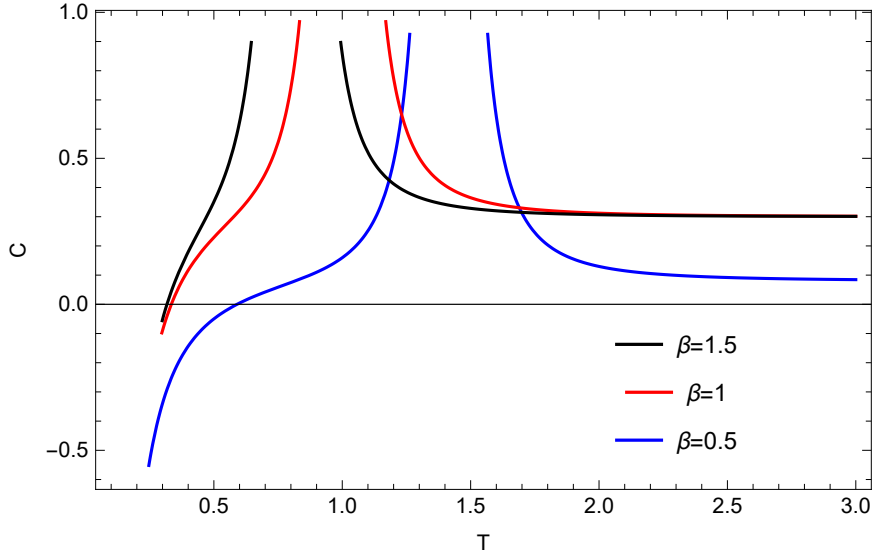}
\end{center}
\vspace{-0.5cm}
\caption{Plot for the specific heat capacity $C$ as a function of temperature $T$ for different values of $\beta$. This is a graphical representation and the other constants are chosen to be unity. Plot not to scale.}
\label{figfth}
\end{figure}

In Fig.(\ref{figfth}) we have plotted the specific heat capacity $C$ against the temperature $T$. It is seen that there is a transition from positive to negative regime in all the trajectories. The temperature at which the transition occurs is the remnant temperature $T_{rem}$ corresponding to $C=0$ and the remnant mass $M_{rem}$.

\vspace{1cm}

Finally, we derive the effective black hole mass via the definition given in
Eqs. () and (). We find 
\begin{equation}
\Delta P=F\Delta t=\frac{2GM}{r^{3}}\Delta t\Delta E\Delta X.
\end{equation}%
Multiplying both side of the above \ inequality by ( $\Delta P$) and using
the Heisenberg uncertainty relationships, we find

\begin{equation}
\left( \Delta P\right) \simeq \sqrt{\hbar ^{2}\frac{2GM}{r^{3}}},
\label{DELATA}
\end{equation}%
next, we consider $\Delta P$ of order $P$, and we substitute (\ref{DELATA})
into (\ref{GUPM}), we obtain the expresson of effective black hole mass as

\begin{equation}
\bar{M}=M\left( 1-2\beta \hbar ^{2}\frac{GM}{r^{3}}\right) .
\end{equation}

\section{Conclusion}
In this work we have explored the effects of higher order GUP on the properties of a black hole. This effects of GUP are inspired from the famous quantum gravity COW experiments and the Einstein-Bohr's photon box thought experiment. Two different types of GUP models inspired from different ideas were considered. They were GUP to all orders in the Planck length model and GUP with minimal length uncertainty and maximal momentum model. For each case we studied the modified de Broglie formula, modification in the gravitational phase shift, the implications of the Einstein-Bohr's photon box experiment, and the modified Hawking temperature and the tidal force of the GUP modified black hole. To study the properties of the modified black hole plots for black hole mass and specific heat capacity were generated. It was seen that the black hole mass gradually decayed with temperature, finally reaching a constant state corresponding to the remnant mass. For the specific heat it was evident that there was a transition from negative to positive values. The temperature value at which the specific heat vanished corresponds to the remnant temperature and the remnant mass. The properties of the GUP modified black hole are quite different from a classical black hole. Due to the effect of GUP the complete evaporation of the black hole due to Hawking radiation is restricted and black hole reaches a remnant stage of inertness.

\section{Acknowledgement}
P.R. acknowledges the Inter University Centre for Astronomy and Astrophysics (IUCAA), Pune, India for granting visiting associateship.

\bigskip \

\end{document}